\documentclass[a4paper]{article}

\usepackage[colorlinks=true,linkcolor=blue,citecolor=blue,urlcolor=blue]{hyperref}

\usepackage[utf8]{inputenc}
\usepackage{amsmath,amssymb,amsthm}
\usepackage{graphicx} % For \rotatebox
\usepackage{multirow} % For multirow cells
\usepackage{caption}
\usepackage{booktabs} 
\usepackage{subcaption}
\graphicspath{{./figures/}}
\usepackage{appendix}
\usepackage{float}
\usepackage{hyperref}
\usepackage{xcolor}
\usepackage{float}  % For Fig/Tables place
\usepackage{hyperref}
\usepackage{xifthen}
\newcommand{\textcite}[2][]{\ifthenelse{\isempty{#1}}{\cite{#2}}{\cite[#1]{#2}}}
\newcommand{\parencite}[2][]{\ifthenelse{\isempty{#1}}{\cite{#2}}{\cite[#1]{#2}}}

\begin{document}

\title{Machine Learning Models for Accurately Predicting Properties of  $\mathrm{CsPbCl}_3$ Perovskite  Quantum Dots}
\author{Mehmet Sıddık Çadırcı$^1$$^*$,  Musa Çadırcı$^2$}
\date{
	\slshape\small 
 $^1$ Faculty of Science, Department of Statistics, Cumhuriyet University, Sivas, Turkey.  \\
    $^2$ Department of Electrical$\And$Electronics Engineering, Duzce University, Düzce, Turkey.  \\[3ex] 
	\normalfont
	%\today 
}
\maketitle$^*$ \textbf{Corresponding author}, \textbf{Email:}msiddikcadirci@cumhuriyet.edu.tr
\begin{abstract}

Perovskite Quantum Dots (PQDs) have a promising future for several applications due to their unique properties. This study investigates the effectiveness of Machine Learning (ML) in predicting the size, absorbance (1S abs) and photoluminescence (PL) properties of  $\mathrm{CsPbCl}_3$ PQDs using synthesizing features as the input dataset. the study employed ML models of Support Vector Regression (SVR), Nearest Neighbour Distance (NND), Random Forest (RF), Gradient Boosting Machine (GBM), Decision Tree (DT) and Deep Learning (DL). Although all models performed highly accurate results, SVR and NND demonstrated the best accurate property prediction by achieving excellent performance on the test and training datasets, with high $\mathrm{R}^2$ and low Root Mean Squared Error (RMSE) and low Mean Absolute Error (MAE) metric values. Given that ML is becoming more superior,  its ability to understand the QDs field could prove invaluable to shape the future of nanomaterials designing.
\end{abstract}

%========================================
\section{INTRODUCTION}\label{sec:intro} 
%========================================

Within computer science, artificial intelligence (AI) refers to the study of designing intelligent machines which can sense the environment around them and take appropriate actions accordingly \cite{russell2016artificial}. Machine learning (ML) is a branch of artificial intelligence that employs algorithms to develop mathematical models from data for solving specific problems directly without applying physical principles that gave birth to the data. This method is especially valuable when the connection between the variables used in the study and the results of the study is not understood. Widespread access to computational power along with the increasing amount of data available for experimentation has resulted in the emergence and application in different areas of science and industry of advanced machine learning models \textcite{liu2020deep, senior2020improved}. Thanks to the capability of evaluating the massive amount of data, Machine learning (ML) is significant for predicting QDs’ properties with a high accuracy. Since QDs’ properties are highly dependent on the size, and composition\cite{gundougdu2021third}, ML algorithms are appropriate tools to handle the data and exhibit well-expressed interactions between the input variables and the resultant properties. Moreover, ML can improve the procedure of synthesizing QDs in order to give out desired characteristics without running more expensive tests and complex simulations which take much time\cite{tao2021nanoparticle}.  Besides, ML can unearth hidden patterns from data that aid scientists in understanding new mechanisms and links in QDs\cite{sanchez2018inverse}\cite{yao2021inverse}. Predicting properties of QDs in diverse conditions using ML is important in materials design for specific applications\cite{tao2021nanoparticle}. 

All inorganic metal halide Perovskite Quantum Dots (PQDs) have shown great promise due to their unique optical and electronic properties. They exhibit size- and composition-dependent properties and are cubic in shape, with the size ranging from approximately 3 nm to ~15 nm. Therefore, PQDs offer band gap tunability over a wide range and property tailoring capabilities. Compared to their counterparts, PQDs have high photoluminescence quantum yield, narrow emission linewidth, higher stability, higher charge mobility, and longer diffusion length \cite{huang2017lead}\cite{yettapu2016terahertz}\cite{wu2015ultrafast}\cite{de2016highly}\cite{wang2015all}.  These properties make them exceptionally critical in a wide range of applications; including solar cells\cite{zhao2019high}, lasers\cite{wang2015all}, LEDs\cite{li2019perovskite}, and medical imaging\cite{ryu2021vivo}. PQDs are produced using the colloidal synthesizing method at high temperatures where the temperature, reaction time, and the amount of substance are significantly vital for the properties of PQDs.  Therefore, to obtain PQDs with the anticipated optical and electronic properties, it is necessary to carefully design the experiment conditions with the full knowledge of every parameter’s effect.  This process is usually carried out by trial and error, which is time-consuming, costly and requires intensive manpower. In this context, ML is a powerful tool to predict the precise conditions of experiments for synthesizing PQDs with the desired properties. It has been shown that ML enables researchers to extract valuable insights from large datasets such as  forecasting a variety of chemical  and
physical features of materials to find intricate mathematical relationships within
empirical data \textcite{lo2018machine}. Different ML algorithms have been applied to the synthesizing conditions of several QDs to predict their sizes including, CdSe\textcite{baum2020machine}, PbSe\textcite{baum2020machine}, ZnSe \textcite{baum2020machine}, InP\cite{nguyen2022predicting} and ZnO\cite{regonia2020predicting} QDs. However,   a comprehensive ML study has not been conducted to predict the optical properties of PQDs yet. 

In this work, we applied several ML algorithms to predict the output properties of $\mathrm{CsPbCl}_3$ PQDs synthesized via the hot-injection method. Based on the predicted photoluminescence, absorption and size properties of $\mathrm{CsPbCl}_3$ PQDs, we compared the performances of ML models of SVR, NND, Deep Learning, Decision Tree, Random Forest, and GBM. 

%========================================
\section{METHODOLOGY}\label{sec:meth}
%========================================

\subsection{Data Description}

We initiated our research by thoroughly analysing existing literature to compile a comprehensive database of hot injection synthesis parameters for $\mathrm{CsPbCl}_3$ PQDs. The data was collected from a total of 59 peer-reviewed articles which are listed in Table S1  in the supporting information section.  
 Once the selected papers were decided, relevant synthesis parameters and the corresponding output properties were extracted manually. The following parameters are considered as the independent input variables to train algorithms; The injection temperature, the source of chloride (Cl), the amount of Cl in millimoles (mmol), the source of lead (Pb), the amount of Pb in mmol, the cesium (Cs) source, the quantity of Cs in mmol, the molar ratio of Cs-to-Pb, and the molar ratio of Cl to Pb.
In addition, the quantities of octadecene (ODE), oleic acid (OA), and oleylamine (OLA) in millilitres (ml), along with the total volume of ligands (OA+OLA) in ml, are also included as input parameters. Furthermore, the ratio of Cl amount to total ligand volume and the amount of Pb to total ligand volume are also utilised as input features. 
The output target parameters are PQDs' size in nanometer (nm), the 1S abs peak in nm and the PL in nm.  We suitably classified the collected data, each variable parameter located in its respective columns, and every outcome in its respective rows. This well-ordered records set improved the model better trained and quicker in data management.

\subsection{Machine Learning Models and Metrics}

 The dataset is separated into training and testing categories according to the hierarchical clustering framework instead of using the same ones repetitively to avoid cases where memorizing or overfitting hinders new information. We evaluated the six regression methods that are suitable for small datasets:  SVR, NND, DL, DT, RF and GBM. All of these algorithms were built using the sklearn library. To guarantee representative samples for testing and training, we utilised both random sampling and stratified sampling techniques. 
We partitioned our data sets into training which contained 80\% examples while testing contains the remaining 20\% examples.  The tuning hyperparameters were performed through Grid search. We evaluated the model's performance by computing $\mathrm{R}^2$, MAE and RMSE metrics. MAE mainly considers outliers and compares datasets, and models with different objectives measured on the scale. A simple way to visualise the model's performance is to look at its MAE value; lower values correspond to higher predicted accuracy. The distance between the predicted actual value and the observed value of the data sample is the best way to interpret RMSE. If RMSE equals zero, then the model is correctly estimating all prices. The coefficient of determination noted as $\mathrm{R}^2$ is a metric that quantifies the degree to which the model accurately represents the data, with values closer to 1 suggesting a higher level of accuracy. 

In data science,  SVR is a line of regression model that is effective in modeling complex relationships within a dataset through the mapping of input data into a higher dimensional space. SVR’s application is of importance especially in dealing with high-dimensional data sets and not-linear relationships. Which can make SVR computationally intensive especially with large datasets
The SVR model was created using the radial basis function (RBF) kernel with the scikit-learn Python module. The hyperparameters were optimised using a grid search technique.

NND is an important concept in spatial analysis and machine learning. More precisely, it plays a significant role in pattern recognition as well as classification algorithms which includes k-Nearest Neighbor (k-NN) method. NND is defined as the shortest length that separates any two points within the dataset from one another. NND has been applied in computational geometry foundational concepts, particle systems theoretical analysis, and statistical estimator convergence analysis, according to \textcite{liitiainen2008bounds}. The Python scikit-learn library was also utilised to implement the NND model.

DT are straightforward but robust models that are simple to understand and illustrate. They can deal with both numerical and categorical data, hence becoming flexible for diverse data sets. Nevertheless, decision trees are susceptible to overfitting specifically as the tree becomes too deep. To train the model, a decision tree model was developed using Python's scikit-learn module.  The model's parameters, including its maximum depth, were changed by applying cross-validation.

RF is a machine learning algorithm that has become famous in recent days \textcite{zhou2018random}. It is considered one of the best machine learning algorithms by many people because: it can handle a thousand variables without compromising the accuracy; it is fast; it is simple to implement; and its prediction accuracy is high \textcite{novita2021br+}.  This algorithm has been referred to as one with high-level prediction performance but requiring less tuning hence regarded as the most appropriate out-of-the-box classification and regression algorithm \textcite{rhodes2023geometry}. The RF model was implemented in Python's scikit-learn module. 500 trees were used to train the model, and cross-validation was used to optimise max\_features, the number of features to take into account at each split.

 GBM is another machine learning technique that is significantly strong because it combines many weak learners. This technique is efficient in many classification tasks \textcite{dong2018grcan}. It has also been identified for its high predictive accuracy and effectiveness when working with complicated interactions in the data. However, it tends to overfit when not well-adjusted as one of the drawbacks of GBM. Python was used to train the GBM model with the scikit-learn library. By implementing cross-validation on critical parameters such as learning rate, number of boosting rounds and max\_depth, they were properly optimized.

DL, particularly neural networks, possess the potential to learn through examples in the same way humans do. These networks do not require specific algorithms and are capable of estimating any nonlinear transformation; hence they can be used to determine inputs/outputs for intricate systems \textcite{zhao2021review}. Nonetheless, there are problems associated with using older model architectures that include a lack of balance within the dataset resulting in memorization rather than generalization by machine learning algorithms themselves as well as redundancy within feature extraction along with ignoring cross-layer characteristic interactions \textcite{ma2023effnet}. We used scikit-learn library for training our RF model in Python.

%========================================
\section{RESULTS AND DISCUSSION}\label{sec:res_dis}
%========================================
The numerical simulations performed in this study have given a more detailed account of how well machine learning techniques could forecast the Size, 1S abs and PL for $\mathrm{CsPbCl}_3$ PQDs. The employed models were    SVR, NND, DT, RF,  GBM and  DL. For training and testing data sets, standard measures like RMSE, MAE and $\mathrm{R}^2$ were used in order to assess the models's performance levels. A series of numerical experiments have given us a full and clear understanding of forecasting $\mathrm{CsPbCl}_3$ PQDs' size, 1S abs and PL output through the indicated ML  models.

%********************
\begin{figure*}[htbp]
\centering\includegraphics[scale=0.33]{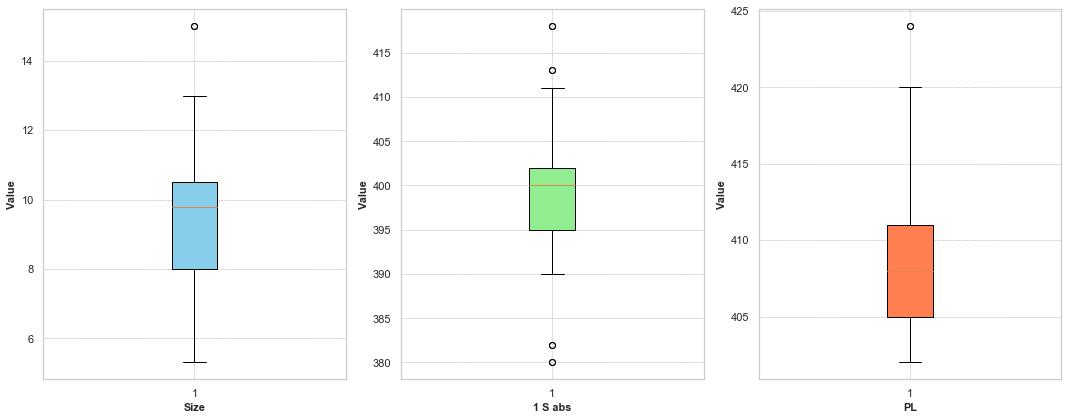}
\caption{Box plots for the  \text{Size}, \text{1S abs}, \text{PL} providing the data distribution, median, quartiles, and potential outliers for each variable.}
\label{fig:Boxplot}
\end{figure*}
%********************

%********************
\begin{figure*}[htbp]
\centering\includegraphics[scale=0.45]{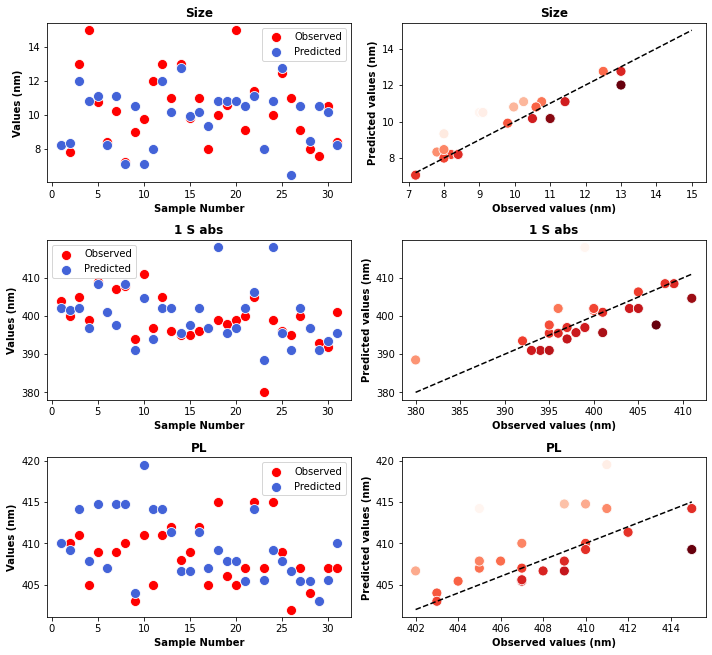}
\caption{Parity plots of predicted vs. observed values for the \text{Size}, \text{1S abs} and \text{PL} outputs of the $\mathrm{CsPbCl}_3$ PQDs using DT regression model}
\label{fig:DT_sample_obs}
\end{figure*}
%********************

 %********************
\begin{figure*}[htbp]
\centering\includegraphics[scale=0.25]{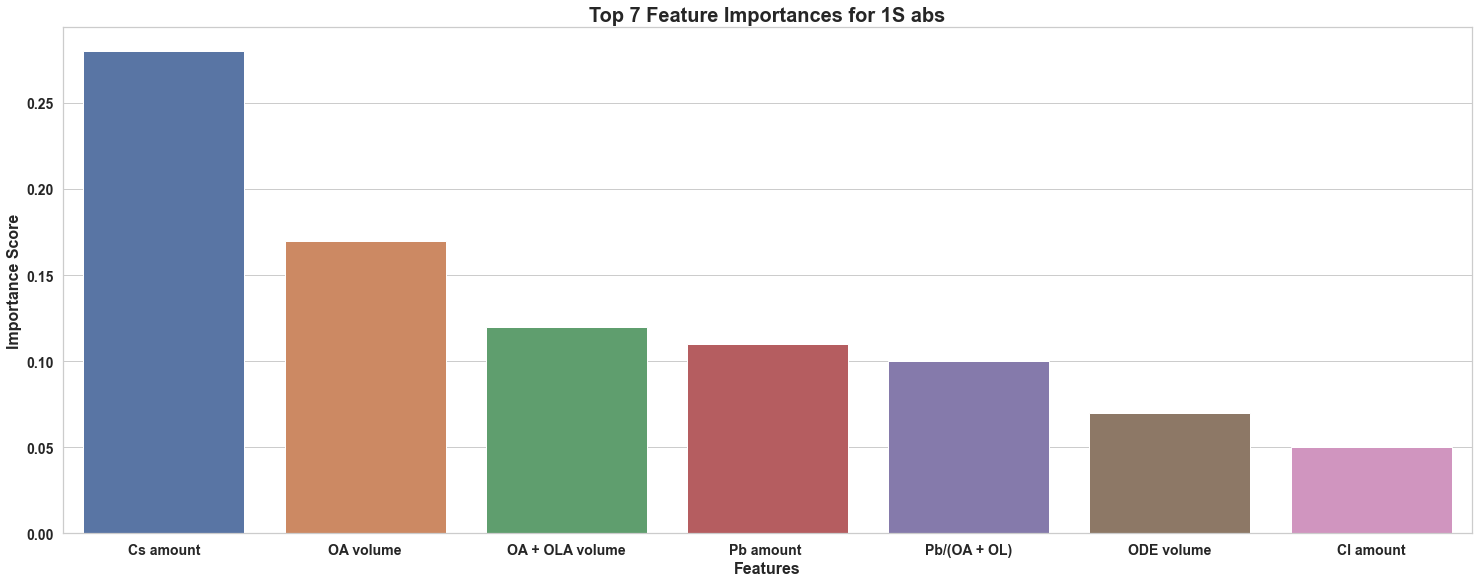}
\caption{Importance of the variables for 1S abs using the RF regression model.}
\label{fig:Importance_of_Absorption}
\end{figure*}
%********************

Initially, to identify the distribution of the data collected from the papers, the outlier data, and data medians, we generated the boxplot for the three output properties as shown in Figure \ref{fig:Boxplot}. It is seen that the median size distribution of PQDs is around 9.5 nm, whereas the 1S abs and PL ranges vary between 395 and 402 nm and 405 and 412 nm, respectively. Next, to compare the actual and predicted processed values using ML algorithms, we conducted scatter graphs of all models for three output properties of  $\mathrm{CsPbCl}_3$ PQDs. Figure \ref{fig:DT_sample_obs}  compares the DT model prediction against actual values of size, 1S abs and PL outputs. Clearly, it shows how accurate the model performs. The predicted and observed values for three output properties of PQDs are almost overlapped in this model. Test data RMSE values of as low as 0.23, 0.19 and 0.16 are respectively obtained for size, 1S abs and PL in this algorithm.  On the other hand, the minimum test data MAE indicators of 0.16, 0.13 and 0.11 are calculated for the properties of size, 1S abs and PL, respectively. Similar Scatter graphs for other employed ML algorithms are given in Figures S1,2,3,4,5  in the supporting information section. These models also yielded similar prediction performances for properties of $\mathrm{CsPbCl}_3$ PQDs. All these findings suggest that ML algorithms are powerful tools for estimating the properties of PQDs accurately.

\clearpage 
%********************
\begin{table}[htbp]
\centering
\caption{Comparison of performance metrics values for  Size,  1S abs, and PL using all ML methods.}
\label{tab:performance_metrics}
\begin{tabular}{|l|l|ccc|ccc|}
\hline
\multirow{2}{*}{\rotatebox[origin=c]{90}{}} & \multirow{2}{*}{\textbf{Model}} & \multicolumn{3}{c|}{\textbf{Train data}} & \multicolumn{3}{c|}{\textbf{Test data}} \\ \cline{3-8}
                  &                      & R$^2$       & RMSE    & MAE     & R$^2$      & RMSE     & MAE      \\ \hline
\multirow{6}{*}{\rotatebox[origin=c]{90}{\textbf{Size}}}     
                  & SVR                  & 0.99      & 0.009  & 0.009 & 0.80     & 0.34   & 0.16   \\
                  & NND                  & 0.99      & 0.012  & 0.008  & 0.62     & 0.47  & 0.30   \\
                  & Deep Learning        & 0.77     & 0.49  & 0.38  & 0.10     & 0.74     & 0.56        \\
                  & Decision Tree        & 0.94      & 0.23  & 0.17  & 0.94     & 0.23   & 0.16   \\
                  & Random Forest        & 0.93     & 0.26  & 0.20  & 0.51     & 0.66   & 0.54   \\
                  & GBM                  & 0.97      & 0.14  & 0.13  & 0.48     & 0.56   & 0.38   \\ \hline

\multirow{6}{*}{\rotatebox[origin=c]{90}{\textbf{1 S abs}}}               
                  & SVR                  & 0.99      & 0.009  & 0.008  & 0.84      & 0.34    & 0.19   \\
                  & NND                  & 0.99      & 0.009  & 0.005  & 0.55      & 0.59    & 0.34   \\
                  & Deep Learning        & 0.66      & 0.59   & 0.39   & 0.44      & 0.66    & 0.49        \\
                  & Decision Tree        & 0.96      & 0.19   & 0.13   & 0.96      & 0.19    & 0.13   \\
                  & Random Forest        & 0.94      & 0.23   & 0.17   & 0.64      & 0.53    & 0.37   \\
                  & GBM                  & 0.98      & 0.11   & 0.09   & 0.66      & 0.51    & 0.30   \\ \hline
\multirow{6}{*}{\rotatebox[origin=c]{90}{\textbf{PL}}}               
                  & SVR                  & 0.99      & 0.009   & 0.009   & 0.66      & 0.58    & 0.28   \\
                  & NND                  & 0.99      & 0.005   & 0.002   & 0.78      & 0.46    & 0.29   \\
                  & Deep Learning        & 0.73      & 0.51    & 0.38    & 0.53      & 0.68    & 0.56      \\
                  & Decision Tree        & 0.97      & 0.16    & 0.11    & 0.97      & 0.16    & 0.11   \\
                  & Random Forest        & 0.94      & 0.23    & 0.16    & 0.70      & 0.54    & 0.39   \\
                  & GBM                  & 0.99      & 0.09    & 0.07    & 0.71      & 0.53    & 0.34   \\ \hline
\end{tabular}
\end{table}
%********************

The input feature importance for predicting 1S abs using the RF algorithm is shown in Figure \ref{fig:Importance_of_Absorption}. The amounts of Cs and OA are seen to be the most significant features, whereas the quantity of Cl and ODE are found to be the less important input features for accurately estimating the indicated property. On the other hand, for size and PL outputs the amounts of Pb and Cs are the most significant input parameters, respectively (See Figures S8 $\And$ S9).

\begin{figure*}[htbp]
\centering\includegraphics[scale=0.4]{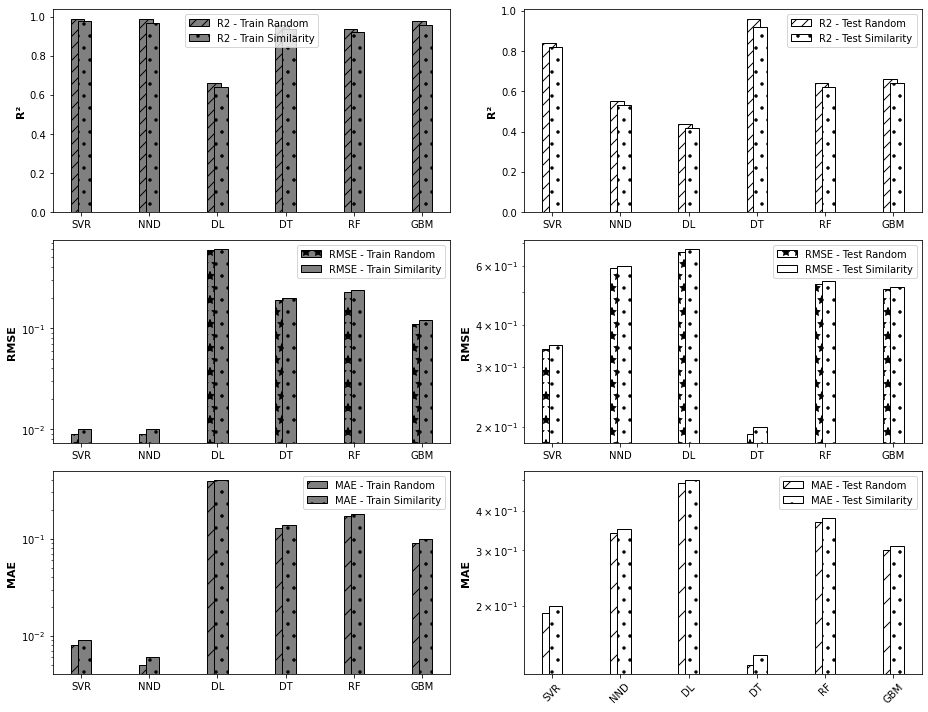}
\caption{Performance metrics for the algorithm trained and tested in \text{1S abs} output, with splitting based on random conditions or to minimize similarity for the training and test data.}
\label{fig:Test_Train_Split_Similarity_S_abs_Nm}
\end{figure*}

To minimize the similarity for the training and test data, we compared  $\mathrm{R}^2$, RMSE, and MAE metrics obtained from test and trained data for 1S abs target output for all employed algorithms, as shown in Figure \ref{fig:Test_Train_Split_Similarity_S_abs_Nm}.  The training data and the testing data appear in different parts of the bar plots. Overall, SVR and NND could perform better than other algorithms for training data, whereas, for test data SVR and DT models outperform their rivals. The metric performance comparison for size and PL outputs of the $\mathrm{CsPbCl}_3$ PQD can be seen in Figures S6 and S7 in the supporting information section.

The Table \ref{tab:performance_metrics}  compiles the metric ($\mathrm{R}^2$, RMSE,
and MAE) performances of training and test data set for three target parameters of $\mathrm{CsPbCl}_3$ PQDs using all ML models. In general, all used ML algorithms utilised in this study provide high accuracy for predicting the target characteristics. 
The SVR model yields the highest $\mathrm{R}^2$ values for three target features in the test data category compared with the other models, which is an indicator of being the most accurate prediction method. In this model, in the trained data category, the RMSE and MAE metrics are found to be as low as 0.009, which gives one of the best accuracy among all prediction models. These observations also agree with the results obtained from Figure S5  for three outputs, where the predicted and observed values are well-correlated. On the other hand, the NND model is also found to be as accurate as the SVR model. It is obvious that the metrics performances of NND model are nearly identical to those of the  SVR model. The well data arrangement in Figure S1 for size, 1S abs, and PL properties of the PQDs, also confirms the accuracy of the NND model.

Conversely,  DL and RF models seem to be the least accurate method for predicting the properties of $\mathrm{CsPbCl}_3$ PQDs. For example, the RMSE and MAE metrics values for predicting test data set for size feature are 0.74 and 0.56, respectively, being 2 times and 3 times lesser those of the  SVR model.  Although the prediction performance of RF and DL for size feature are inferior to others employed in this study, their performances are marginally better than that used for different QDs in the literature \textcite{baum2020machine}. The metrics values for all target parameters obtained from GBM and DT models demonstrate a moderate performance between the SVR-NND and DL-RF algorithm couples. These two models showed better prediction performance when used for PL output estimation.

\begin{figure*}[htbp]
\centering\includegraphics[scale=0.5]{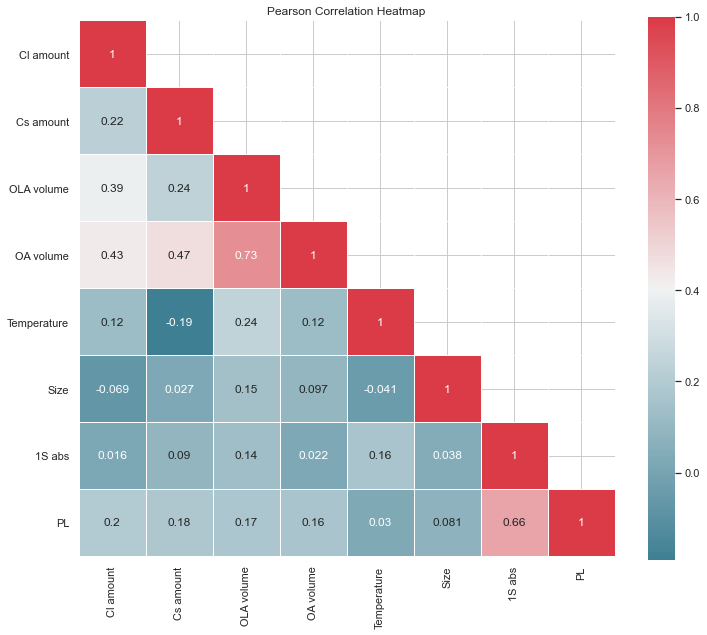}
\caption{Pearson correlation shows a strong correlation between the three output targets}
\label{fig:Pearson_Correlation_Heatmap}
\end{figure*}

A Pearson correlation heatmap is a graphical representation that effectively communicates the Pearson correlation coefficients (ranging between -1 to +1) among variables within a dataset. The heatmap displays a colour-coded matrix, with warmer colours indicating stronger positive correlations, cooler colours representing stronger negative correlations, and neutral colours signifying no link between variables \textcite{rainey2023experimental}.  Figure \ref{fig:Pearson_Correlation_Heatmap} shows the Pearson correlation heatmap to demonstrate the correlations between the input and output parameters dataset of   $\mathrm{CsPbCl}_3$ PQDs.  1S abs and PL have a positive correlation of 0.66, whereas, the correlations between size and PL and size and 1S abs are considerably inferior.

\clearpage 
%********************
%********************

%********************
%********************

From synthesis parameters, results have shown that machine learning can be used to accurately predict the properties of $\mathrm{CsPbCl}_3$ quantum dots. In fact, the most efficient ones were SVR and RF models which performed well in their predictions and also provided some guidance in understanding the parameters affecting the properties of quantum dots. The importance of growth temperature was underlined by RF because these variables have a direct effect on quantum dots characteristics and hence require precise control measures. Another finding was that; NND and DT models though fast at producing easily understood outputs might need further optimization if they are to match other sophisticated models like GBM or DL. Boxplots and correlation analyses serve as an additional level of validation for ensuring predictive model stability. In general, this paper demonstrates how ML can help in the control and synthesis of QDs implying a route to more efficient experimental designs.

%\newpage
\clearpage 
%========================================
\section{CONCLUSION}\label{sec:conc}
%========================================

This study aimed to predict the size, 1S abs and PL properties of $\mathrm{CsPbCl}_3$ PQDs by comparing the performances of ML algorithms of SVR, NND, GBM, RF, DT and DL. Generally, nearly all models succeeded in promising outcomes in predicting the outputs of PQDs. Among them,  SVR and NND models indicated the best performance as they make accurate predictions and give insights into factors that affect QD properties. The SVR and NND ML models demonstrate RMSE metric values of 0.009 and 0.012 for the train data, and 0.34 and 0.47 for the test data, respectively. These findings are close to actual data, which indicate that the employed ML models have the capability of predicting properties of $\mathrm{CsPbCl}_3$ PQDs with high accuracy. For the future direction, the results suggest that the progress of ML can significantly contribute to the progress of QDs design, resulting in more tailored QDs with specific properties.

%========================================

%========================================

%========================================
\section*{Supporting Information}
Electronic supporting information (ESI) is accessible: Additional details about the similarity between the training and test data, the significance of the variables in terms of feature importance, and the comparison of predicted and observed outputs of the $\mathrm{CsPbCl}_3$ PQDs and compound databases.
%========================================
 
%========================================
\section*{Acknowledgements}\label{sec:acknow}
%========================================

%%%%%%% References %%%%%%%%%%%%%%%%%%%%%
% for references with bibtex
\clearpage

\bibliographystyle{plain}  
\bibliography{bibliog}

\newpage
%%%%%%%%%%%%%%%%%%%%%%%%%%%%%%%%%%%%%
%========================================
\section*{Supporting Information}\label{sec:supp}
%========================================

\renewcommand{\thefigure}{S\arabic{figure}}
\renewcommand{\thetable}{S\arabic{table}}

\setcounter{figure}{0}
\setcounter{table}{0}

%********************
\begin{figure*}[htbp]
\centering\includegraphics[scale=0.4]{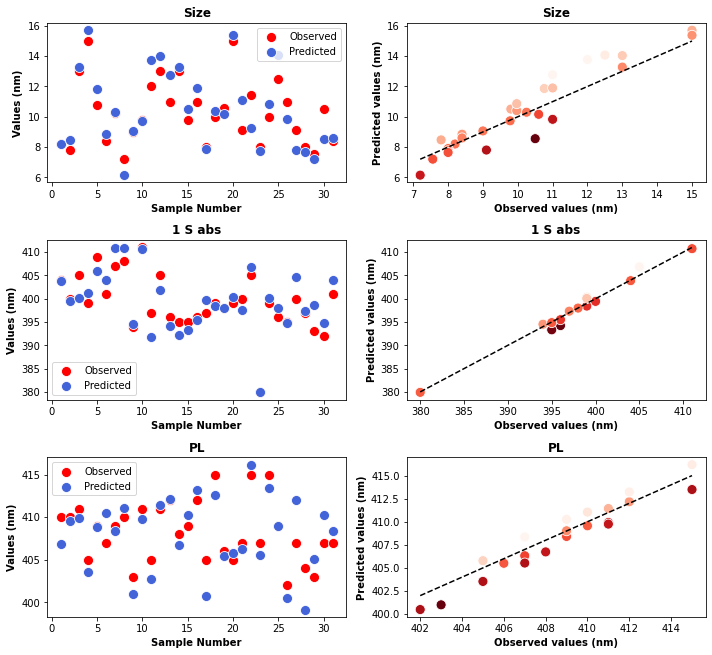}
\caption{Predicted vs. observed plot of the NND   regression model for the \text{Size}, \text{1S abs} and \text{PL}  outputs of the $\mathrm{CsPbCl}_3$ PQDs.}
\label{fig:NND_sample_obs_val}
\end{figure*}
%********************

%********************
\begin{figure*}[htbp]
\centering\includegraphics[scale=0.4]{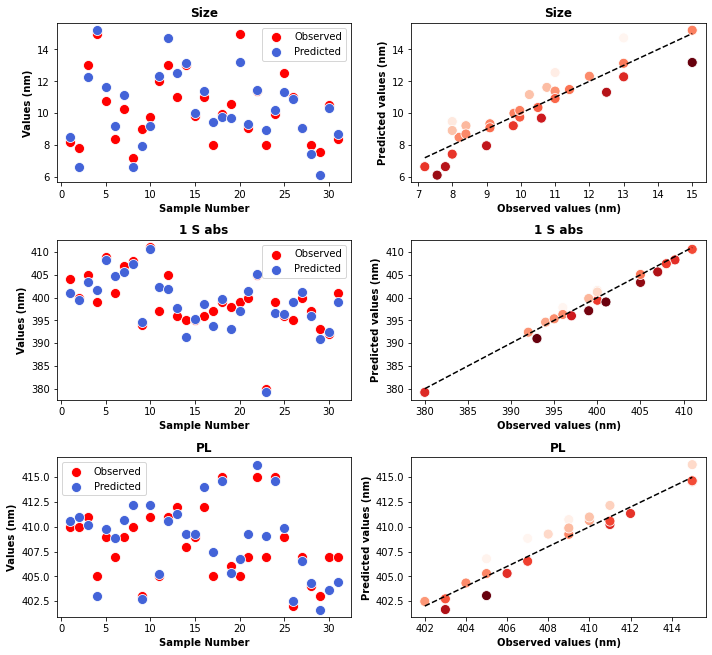}
\caption{Predicted vs. observed plot of the RF regression model for the \text{Size}, \text{1S abs} and \text{PL} outputs of the $\mathrm{CsPbCl}_3$ PQDs.}
\label{fig:random_forest_samp_obs}
\end{figure*}
%********************

%********************
\begin{figure*}[htbp]
\centering\includegraphics[scale=0.4]{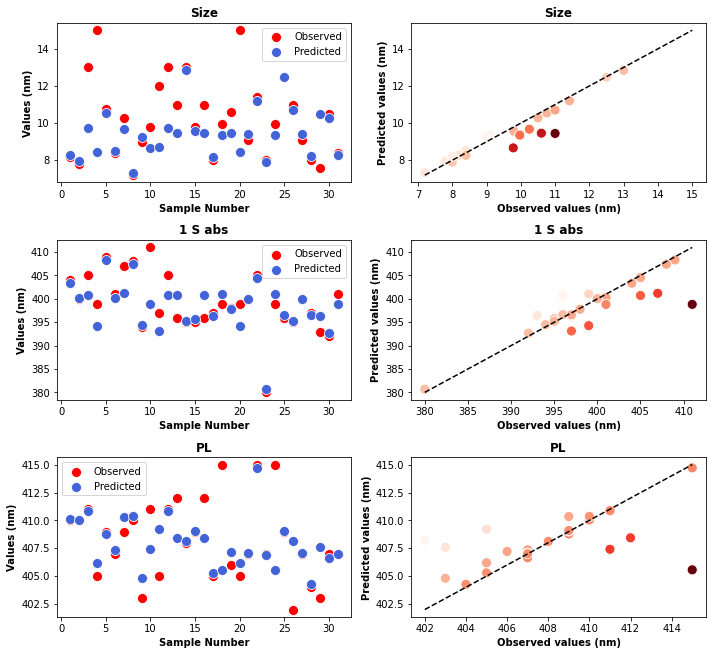}
\caption{Predicted vs. observed plot of the GBM regression model for the \text{Size}, \text{1S abs} and \text{PL} outputs of the $\mathrm{CsPbCl}_3$ PQDs.}
\label{fig:GBM_multi}
\end{figure*}
%********************

%********************
\begin{figure*}[htbp]
\centering\includegraphics[scale=0.45]{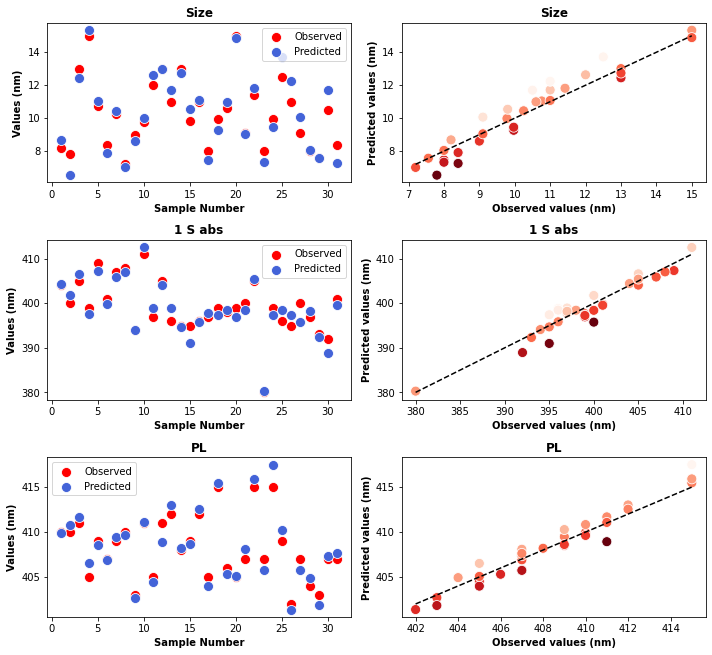}
\caption{Predicted vs. observed plot of the DL regression model for the \text{Size}, \text{1S abs} and \text{PL} outputs of the $\mathrm{CsPbCl}_3$ PQDs.}
\label{fig:DL_samp_obs}
\end{figure*}
%********************
%********************

%********************
\begin{figure*}[htbp]
\centering\includegraphics[scale=0.4]{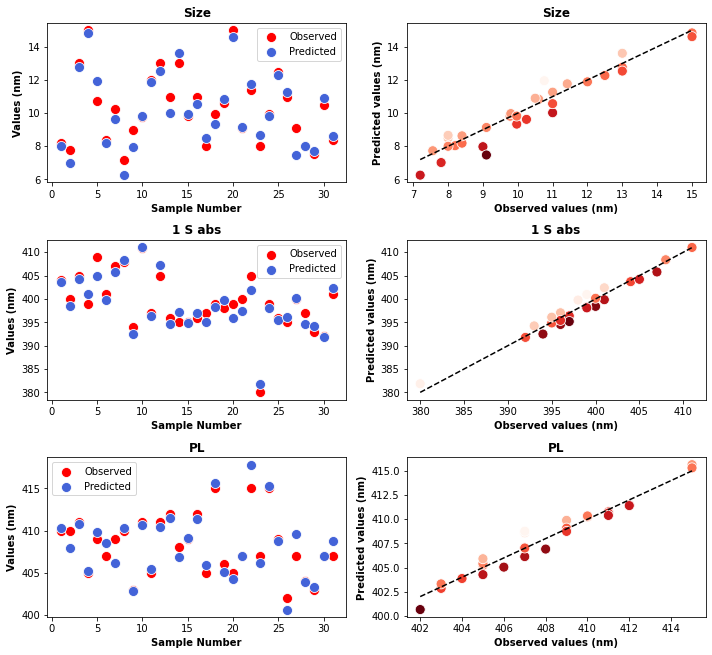}
\caption{Predicted vs. observed plot of the SVR model for the \text{Size}, \text{1S abs} and \text{PL} outputs of the $\mathrm{CsPbCl}_3$ PQDs.}
\label{fig:SVR_samp_Obs_val}
\end{figure*}
%********************

%********************
\begin{figure*}[htbp]
\centering\includegraphics[scale=0.4]{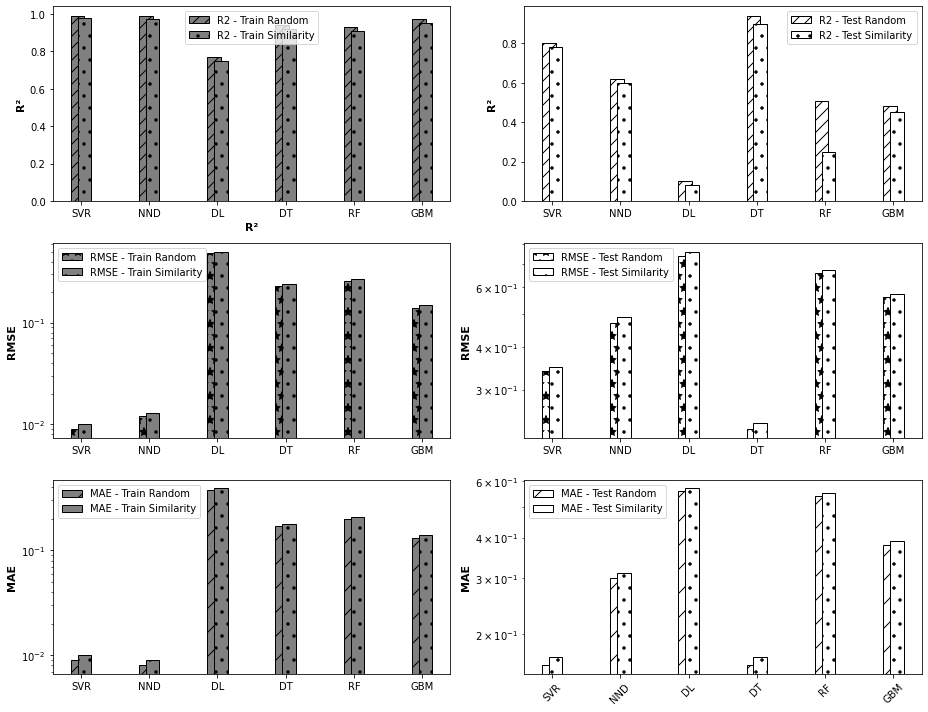}
\caption{Performance metrics for the the algorithm trained and tested in \text{Size} output, with splitting based on random conditions or to minimize similarity for the training and test data.}
\label{fig:Test_Train_Split_Similarity_SIZE_NM_Suppl}
\end{figure*}
%********************

%********************
\begin{figure*}[htbp]
\centering\includegraphics[scale=0.4]{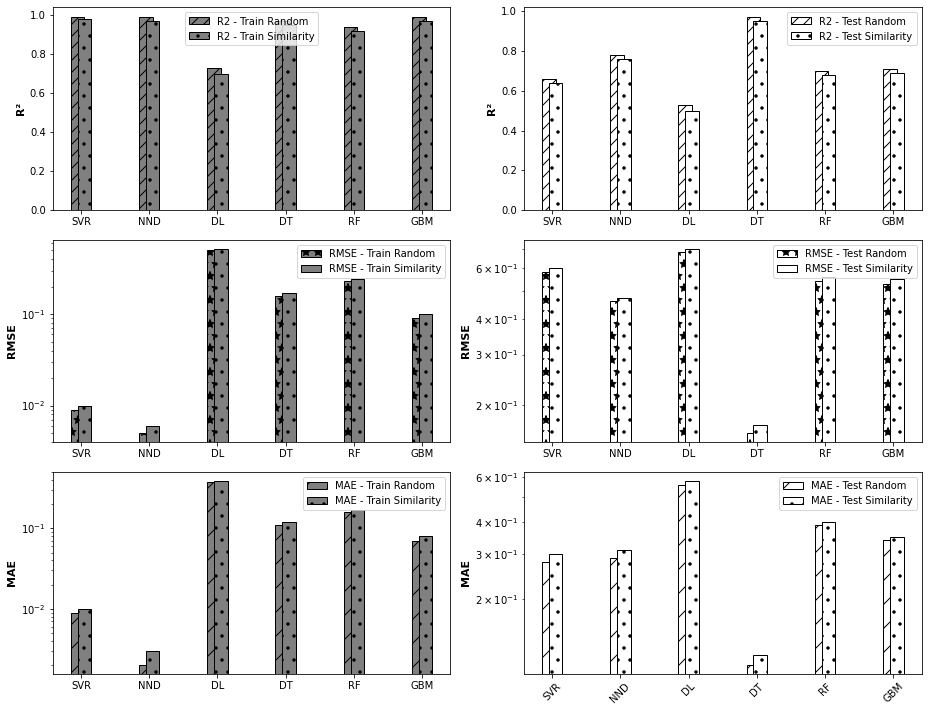}
\caption{Performance metrics for the algorithm trained and tested in \text{PL} output, with splitting based on random conditions or to minimize similarity for the training and test data.}
\label{fig:Test_Train_Split_Similarity_PL_Suppl}
\end{figure*}
%********************

%********************
\begin{figure*}[htbp]
\centering\includegraphics[scale=0.27]{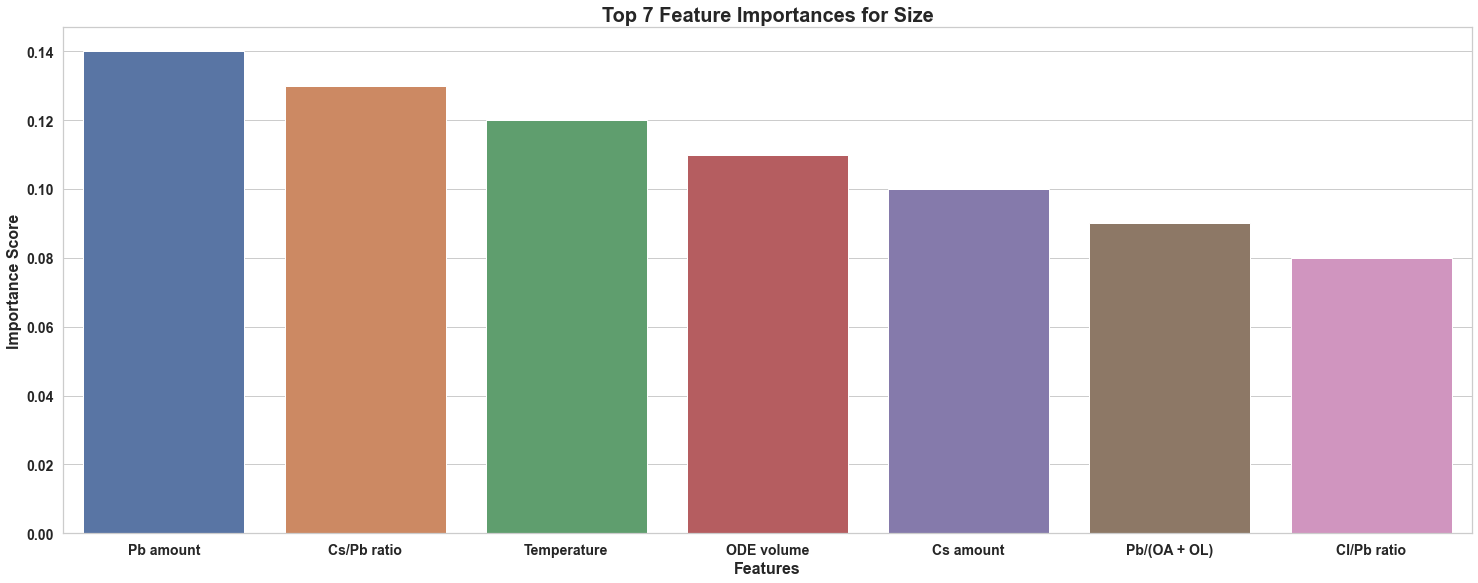}
\caption{Importance of the variables for Size  using the RF regression model.}
\label{fig:Importance_Size_nm}
\end{figure*}
%********************

%********************
\begin{figure*}[htbp]
\centering\includegraphics[scale=0.27]{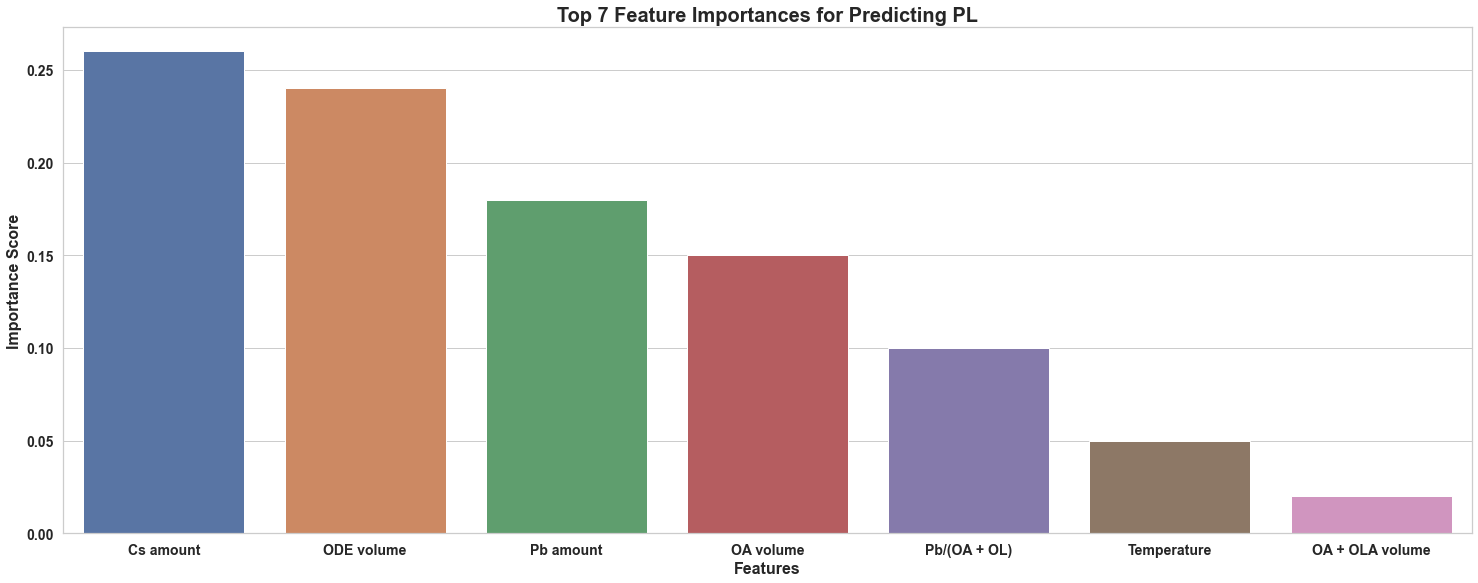}
\caption{Importance of the variables for PL  using the RF regression model.}
\label{fig:Importance_of_PL}
\end{figure*}
%********************

%***************************************************************
\begin{table}[htbp]
\centering
\caption{DOIs are utilised to consolidate the databases.}
\label{tab:S1}
\begin{tabular}{|c|l|c|l|}
\hline
\multicolumn{4}{|c|}{$\mathrm{CsPbCl}_3$ Database DOIs} \\ \hline
01. & 10.1002/adfm.202100930              & 02. & 10.1016/j.jlumin.2021.118658 \\ \hline
03. & 10.1021/acsenergylett.8b01441       & 04. &  10.1021/acsnano.8b07850 \\ \hline
05. & 10.1016/j.solener.2020.05.070         & 06. & 10.1016/j.optmat.2022.113362 \\ \hline
07. & 10.1016/j.materresbull.2018.12.004    & 08. & 10.1007/s00339-022-05962-7   \\ \hline
09. & 10.1021/acsenergylett.9b02678         & 10. & 10.1021/acs.jpclett.8b03047   \\ \hline
11. & 10.1021/acsmaterialslett.9b00101      & 12. & 10.1021/acs.chemmater.9b05082 \\\hline
13. & 10.1016/j.matchemphys.2021.125479     & 14. & 10.1039/D0RA09043C    \\ \hline
15. & 10.1021/acs.nanolett.6b02772         & 16. & 10.1021/jacs.6b08085    \\ \hline
17. & 10.1021/acs.jpclett.1c02416        & 18. & 10.1021/acs.jpcc.8b06579   \\  \hline
19. & 10.1021/acs.chemmater.8b02157        & 20. & 10.1021/acs.jpcc.8b11906    \\ \hline
21. & 10.1021/acs.jchemed.8b00735        & 22. &  10.1021/acsenergylett.8b01909   \\ \hline
23. & 10.1021/acsanm.0c01254         & 24. &  10.1021/acs.jpcc.8b12035   \\ \hline
25. & 10.1039/C7RA06597C         & 26. &  acsanm.1c01623  \\ \hline
27. & 10.1021/acsenergylett.7b00375    & 28. & 10.1039/C8NR10439E   \\ \hline
29. & 10.1016/j.nanoen.2018.06.073         & 30. & 10.1002/asia.202200478     \\ \hline
31. & 10.1016/j.nanoen.2020.105163         & 32. & 10.1021/acs.jpcc.1c04335   \\  \hline
33. & 10.1021/acs.jpcc.7b06929         & 34. & 10.1038/srep45906   \\ \hline
35. & 10.1039/C9RA05685H          & 36. &  10.1016/j.jallcom.2019.02.032   \\ \hline
37. & 10.1039/C9RA07069A         & 38. & 10.1021/acs.nanolett.5b02404  \\ \hline
39. & 10.1021/acs.jpcc.6b12828          & 40. & 10.1021/acs.chemmater.9b05082   \\ \hline
41. & 10.1016/j.materresbull.2020.110907         & 42. & 10.1039/c8tc03957g   \\ \hline
43. & 10.1039/C8TC03139H         & 44. & 10.1063/1.5127887    \\ \hline
45. & 10.1039/c7nr08136g         & 46. & 10.1088/2632-959X/abcf8e   \\  \hline
47. & 10.1039/d1nr04455a         & 48. &  10.1016/j.ceramint.2022.07.310  \\ \hline
49. & 10.1021/acsanm.3c01960         & 50. & 10.1039/c7nr06745c   \\ \hline
51. & 10.1021/acs.jpcc.1c06995         & 52. & 10.1021/acs.chemmater.0c01325   \\ \hline
53. & 10.1021/acs.jpclett.9b03831         & 54. & 10.1021/acsenergylett.0c00931   \\ \hline
55. & 10.1021/acsami.8b14046         & 56. &  10.1021/acs.nanolett.7b04575  \\ \hline
57. & 10.1016/j.ceramint.2022.09.075         & 58. & 10.1021/acs.jpclett.1c00017    \\ \hline
59. &  10.1016/j.jallcom.2021.161505        & 60. &    \\  \hline
\end{tabular}
\end{table}
%***************************************************************

%%%%%%%%%%%%%%%%%%%%%%%%%%%%%%%%%%%%%%%%%
\end{document}